\documentclass[]{aa}
\usepackage{natbib}
\usepackage{lscape,graphicx}
\usepackage{epsf}
\usepackage{subfig}
\usepackage{txfonts}
\bibpunct{(}{)}{,}{a}{}{,}
\normalfont

\def\ergs {erg\,s$^{-1}$}
\def\ergscm2 {erg\,s$^{-1}$cm$^{-2}$}

\def\uu {4U\,0142+614\,}
\def\ee {1E\,1048.1--5937\,}
\def\kes {1E\,1841--045\,}

\def\rxs {1RXS\,J1708--4009\,}
\def\ea {1E\,2259+586\,}
\def\xte {XTE\,J1810--197\,}

\def\sgra{SGR\,1806--20\,}
\def\sgrb{SGR\,1900+14\,}

\def\SPITZ{{\it Spitzer}\,} 
\def\integ{{\it INTEGRAL}\,}

\begin{document}

\title{Adaptive optics near-infrared observations of magnetars}

\author{V. Testa$^{1}$, N. Rea$^{2,3}$, R.~P. Mignani$^{4}$, G.L. Israel$^{1}$,
R. Perna$^{5}$,  S. Chaty$^{6}$, L. Stella$^{1}$, S. Covino$^{7}$, \\  R.
Turolla$^{8,4}$, S. Zane$^{4}$,  G. Lo Curto$^{9}$, S. Campana$^{7}$, G.
Marconi$^{9}$, S. Mereghetti$^{10}$}

\institute{INAF - Osservatorio Astronomico di Roma, Via Frascati 33, 00040 Monte
Porzio Catone, Italy 
\and University of Amsterdam, Astronomical Institute ``Anton Pannekoek'',
Kruislaan 403, 1098~SJ, The Netherlands 
\and SRON Netherlands Institute for Space Research, Sorbonnelaan, 2, 3584~CA,
Utrecht, The Netherlands 
\and MSSL, University College London, Holmbury St. Mary, Dorking Surrey RH5 6NT,
UK 
\and JILA and Department of Astrophysical and Planetary Sciences, University of
Colorado, 440 UCB, Boulder, CO, 80309, USA 
\and Laboratoire AIM, CEA/DSM --CNRS-- Universit\'e Paris Diderot,
DAPNIA/Service d'Astrophysique, B\^at.709, CEA--Saclay, FR-91191 Gif-sur-Yvette,
France 
\and INAF - Osservatorio Astronomico di Brera, via Bianchi 46, 23807 Merate
(Lc), Italy  
\and University of Padua, Physics Department, via Marzolo 8, 35131, Padova,
Italy 
\and ESO - European Southern Observatory, Alonso de Cordova 3107, Vitacura,
Santiago, Chile 
\and INAF - Istituto di Astrofisica Spaziale e Fisica Cosmica `G.Occhialini',
via Bassini 15, 20133, Milano, Italy}

\offprints{V. Testa: testa@mporzio.astro.it}

\date{Received / Accepted }

\abstract {We report on near-infrared (IR) observations of the three 
anomalous X-ray pulsars  \xte, \rxs and \kes, and the soft
gamma-ray repeater \sgrb, taken with the ESO-VLT, the Gemini, and the
CFHT telescopes.}  {This work is aimed at identifying and/or
confirming the IR counterparts of these magnetars, as well as at
measuring their possible IR variability.}  {In order to perform
photometry of objects as faint as K$_s \sim 20$, we have used data
taken with the largest telescopes, equipped with the most advanced IR
detectors and in most of the cases with Adaptive Optics devices. The
latter are critical to achieve the sharp spatial accuracy required to
pinpoint faint objects in crowded fields.}  {We confirm with high
confidence the identification of the IR counterpart to \xte, and its
IR variability. For \kes\, and \sgrb\, we propose two candidate IR
counterparts based on the detection of IR variability. For
\rxs\, we show that none of the potential counterparts within the source
X-ray error circle can be yet convincingly associated with this AXP.}
{The IR variability of the AXP \xte\, does not follow the same monotonic decrease of its post-outburst X-ray emission.
Instead, the IR variability appears more similar to the one observed in radio band, although simultaneous IR and radio observations are crucial to draw any conclusion in this respect. For \kes\, and \sgrb\, follow-up
observations are needed to confirm our proposed candidates with higher confidence. }


\authorrunning{Testa et al.}
\titlerunning{Adaptive optics near-IR observations of magnetars}

\maketitle

\section{Introduction}

In recent years, the study of isolated neutron stars has become one of
the most challenging research areas in high-energy astrophysics,
largely as a result of the discovery of several new classes of sources
besides the classical, well-studied radio pulsars.  The most extreme
objects are the so-called ``magnetars''.  This class comprises the
Anomalous X--ray Pulsars (AXPs) and the Soft Gamma--ray Repeaters
(SGRs), observationally very similar in many respects.  They all are
slow X--ray pulsars with periods in a narrow range  ($P$=
2--12\,s), relatively large period derivatives ($\dot P =
10^{-13}-10^{-10}$s\,s$^{-1}$), spin-down ages of $10^3-10^4$\,yr, and
magnetic fields, as inferred from the classical magnetic dipole
spin-down formula\footnote{$B\sim3\times10^{19}\sqrt{P\dot{P}}$~G},
of $10^{14}-10^{15}$\,G, larger than the electron quantum critical field
($B_{cr}\simeq4.4\times10^{13}$\,G) above which quantum effects become
crucial.  AXPs and SGRs are strong X-ray emitters, with X-ray
luminosities of about $10^{34}-10^{36}$\ergs. Their 0.1--10\,keV
persistent emission has relatively soft spectra usually modeled by an
absorbed blackbody (kT$\sim$\,0.2--0.6) plus a power-law
($\Gamma\sim$\,2--4; for a review see e.g. Woods \& Thompson 2006).
Their X-ray energy output is much larger than the rotational energy
loss and implies that these sources are not rotationally powered,
at variance with most young isolated neutron stars. Rather, AXPs and SGRs are believed to be powered by
the neutron star ultra-strong magnetic field (Duncan \& Thompson 1992; Thompson \& Duncan 1995).  In the ''magnetar'' model,
crustal deformations, driven by magnetic stresses imparted to the
crust by the strong internal toroidal magnetic field, are responsible
for the observed activity, X/$\gamma$-ray bursts and giant flares.

\begin{table*}
\centering
{\scriptsize
\begin{tabular}{cccccccc}
\hline
\hline
Telescope/Instrument & Target & Date (YY.MM.DD) & MJD      & Filter & Exposure
Time (s) &  FWHM($^{\prime\prime}$) & airmass   \\ \hline\hline 
VLT/NACO       & \xte\  & 2003.10.08 & 52920  & H      & 2520 &  0.189 & 1.17   
 \\ 
               &        & 2003.10.08 & 52920  & Ks     & 2520 &  0.243 & 1.49   
 \\ 
               &        & 2004.03.14 & 53078  & H      & 1080 &  0.108 & 1.47   
 \\ 
               &        & 2004.03.14 & 53078  & Ks     & 1200 &  0.094 & 1.20   
 \\ 
               &        & 2004.03.14 & 53078  & J      & 1080 &  0.081 & 1.13   
 \\ 
               &        & 2004.09.09 & 53257  & J      & 1800 &  0.170 & 1.01   
 \\ 
               &        & 2004.09.09 & 53257  & H      & 3600 &  0.140 & 1.08   
 \\ 
               &        & 2004.09.13 & 53261  & Ks     & 3600 &  0.099 & 1.01   
 \\ \hline
               & \rxs\  & 2003.05.20 & 52779  & Ks     & 2400 &  0.083 & 1.04   
 \\ 
               &        & 2003.06.19 & 52809  & H      & 2400 &  0.122 & 1.40   
 \\ 
               &        & 2003.06.20 & 52810  & J      & 2400 &  0.122 & 1.04   
 \\
               &        & 2004.03.25 & 53089  & L'     & 2100 &  0.119 & 1.04   
 \\ \hline
               & \sgrb\ & 2006.04.01 & 53826  & Ks     & 1560 &  0.125 & 1.43   
 \\ 
               &        & 2006.07.21 & 53937  & Ks     & 2040 &  0.111 & 1.21   
 \\ \hline\hline 
Gemini/NIRI    & \xte\  & 2003.09.18 & 52900  & Ks     & 285  &  0.397 & 1.62   
 \\ 
               &        & 2003.09.18 & 52900  & H      & 285  &  0.397 & 1.99   
 \\ 
               &        & 2003.09.18 & 52900  & J      & 285  &  0.397 & 1.78   
 \\ 
               &        & 2004.06.10 & 53166  & Ks     & 510  &  0.631 & 1.32   
 \\ 
               &        & 2004.06.10 & 53166  & H      & 510  &  0.526 & 1.30   
 \\ 
               &        & 2004.06.10 & 53166  & J      & 510  &  0.655 & 1.32   
 \\
               &        & 2004.07.28 & 53214  & Ks     & 510  &  0.468 & 1.39   
 \\  
               &        & 2004.07.28 & 53214  & H      & 510  &  0.526 & 1.53   
 \\ \hline \hline  
CFHT/AOBIR     & \kes\  & 2002.08.17 & 52503  & K'     & 1260 &  0.137 & 1.11   
 \\
               &        & 2002.08.17 & 52503  & H      & 1260 &  0.161 & 1.31   
 \\ \hline
\end{tabular}}
\caption{Observations summary. }
\label{tab:obs}
\end{table*}

\begin{figure*}
\centering
\hbox{
\includegraphics[width=0.49\textwidth]{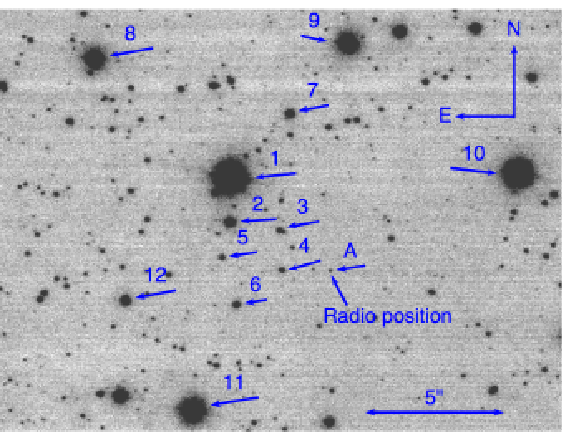}
\includegraphics[width=7.6cm]{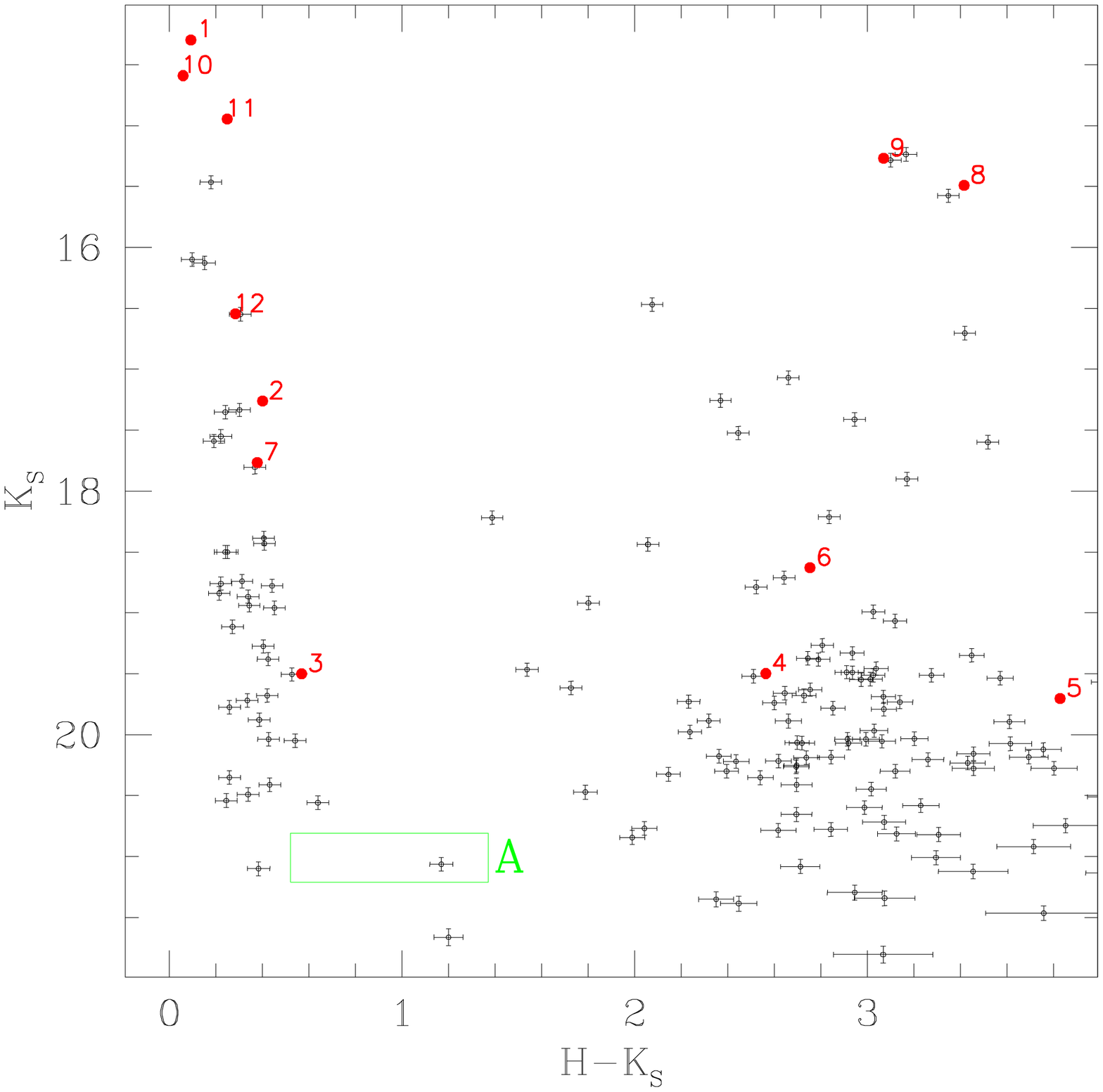}}

 \caption{\xte: {\em Left panel}: VLT/NACO K-band image (March
 2004). The radio position \citep{ca07a} is marked together with
 its candidate counterpart (object A).  
Twelve nearby objects, some of which used as a reference
for relative photometry, are also marked in the figure.  {\em Right
panel:} K$_S$ vs.H-K$_S$ colour-magnitude diagram of the sources detected in
the NACO field of view (filled circles). Object A is marked with a
rectangle indicating the range in colour and magnitude covered throughout
the epochs, while some of the nearby objects marked in the map and
listed in Tab.\,\ref{tab:xteJ} are plotted in red and labeled. 
Magnitudes and colours for all the sources but object 'A' 
have been averaged over all the observations.}

\label{fig:MapXTEJ1810}
\label{fig:CmdXTEJ1810}
\end{figure*}

Until  not long  ago, the  persistent emission  of AXPs  and  SGRs was
detected only in the soft  X-ray range.  However, in the last
few  years the  availability of  more sensitive  optical  and IR
telescopes, as well as  of $\gamma$-ray satellites, opened new windows
in the study of AXPs  and SGRs, unveiling their multi-band properties.
In  particular, observations  from  ESO, Gemini,  CFHT  and Keck  (see
Israel  et al.  2004a for  a  review) led  to the  discovery of  faint
(K$_s\sim$20),  and  in  some  cases  variable, IR  counterparts to  five out  of seven
confirmed AXPs (one of which was also detected in the optical band by Hulleman
et   al.~(2000)),   and   to   one   (out  of   four)   confirmed   SGR
\citep{is05,ko05}.   Furthermore,  deep  \integ\ observations  in  the
20--200\,keV  band  revealed  that  most of  these  highly  magnetized
sources are also hard X-ray emitters \citep{ku04,ku06}.
The recent discovery of pulsed radio emission from the transient
magnetars \xte\ \citep{ha05,ca06} and 1E 1547.0$-$5408 (Camilo et al.
2007a), the only cases so far
\citep{bu06,bu07}, is intriguing since  radio emission was believed to
be  quenched  in magnetic  fields  above  the  quantum critical  limit
(Baring \& Harding 1998).

Despite the more complete observational picture we have now 
for AXPs and SGRs, the physical processes at the basis of their emission 
in the different bands are not fully understood as yet.
This is particularly true as far as the optical and IR emission of these 
sources is concerned. The extrapolation of the canonical blackbody plus power-law, used to model
the soft X-ray emission of these sources, largely over-predicts their
optical and IR emission. Recently, new attempts have been made at
modeling the multi-band spectra of magnetars to overcome this problem.
In particular, \cite{re07a} showed that AXP spectra can be
fitted by a resonant cyclotron scattering model or two log-parabolic
functions, which do not over-predict their dim optical and IR magnitudes. 
However, despite solving this issue, these new spectral
models are still far from giving any physical interpretation for the
optical and IR emission.

Some possibilities are that either the magnetars optical/IR emission arise from the star surface/magnetosphere (Beloborodov \& Thompson 2006), or from the reprocessing of the X-ray emission via a fossil disk around the neutron star (Chatterjee et al. 2000, Perna et al. 2000).  Indeed, the
\SPITZ\  mid-IR  detection of the AXP  \uu\,\citep{wa06} has been
interpreted as the  first evidence of a passive  fossil disk around an
AXP, believed to emit through reprocessing of the X-ray radiation from
the magnetar.

In  this paper we  report on  Gemini and  ESO-VLT observations  of the
transient AXP \xte,  on ESO-VLT observations of the  AXP \rxs\, and of
\sgrb,  and CFHT  observations of  the AXP  \kes  (\S\,\ref{obs}). The
results  are  presented and  discussed  in  the  context of  different
emission  scenarios,  and  compared  with  the  optical  and  IR
properties of other magnetars (see \S\,\ref{discussion}).

\section{Observations and data reduction}
\label{obs}
\subsection{Observations} 

Observations were  performed with the 8\,m Very  Large Telescope (VLT)
and the 3.5m Canada-France-Hawaii Telescope (CFHT), both equipped with
adaptive  optics  (AO) and  near-IR  cameras.   In  addition, we  used
archival observations  performed with the 8\,m  Gemini telescope, also
equipped  with an  IR camera  but without  AO,  that were
offered  starting  from 2004  only. A  summary  of  the  telescopes  and
instruments used, as well as the observations logs, is reported in
Tab.\,\ref{tab:obs}.  In  all cases  images were obtained  through the
standard  observing  technique commonly  applied  to IR  arrays,
i.e. by stacking  repeated exposures along each node  of a pre-defined
dithering  pattern, with  each  node  jittered by  a  few arcsec  with
respect to the  previous one. This allows to obtain  a set of dithered
exposures  that are  used  to produce  sky-background  images.

VLT/Yepun observations of  \xte, \rxs\ and \sgrb\ were  carried out in
 different observing runs (see Tab.\, \ref{tab:obs}) using NAos COnica
 (NACO)\footnote{www.eso.org/instruments/naco}  with  the  S27  camera
 (0\farcs027  pixel  size,  $28^{\prime\prime}\times28^{\prime\prime}$
 field of view). Gemini-North observations of \xte\ were carried out
using the Near-InfraRed Imager
(NIRI)\footnote{www.gemini.edu/sciops/instruments/niri} with the f/6 camera
(0\farcs117 pixel scale, $120^{\prime\prime} \times
120^{\prime\prime}$ field of view).
\kes\ was observed at the CFHT using  
the Adaptive Optics Bonnet (AOB)\footnote{www.cfht.hawaii.edu/Instruments}
InfraRed
camera (0\farcs035 pixel size, $36^{\prime\prime}\times
36^{\prime\prime}$ field of view).

\begin{figure}
\includegraphics[width=3cm,height=7.3cm,angle=270]{xrayflux.ps} 
\includegraphics[width=8.2cm,height=8.3cm]{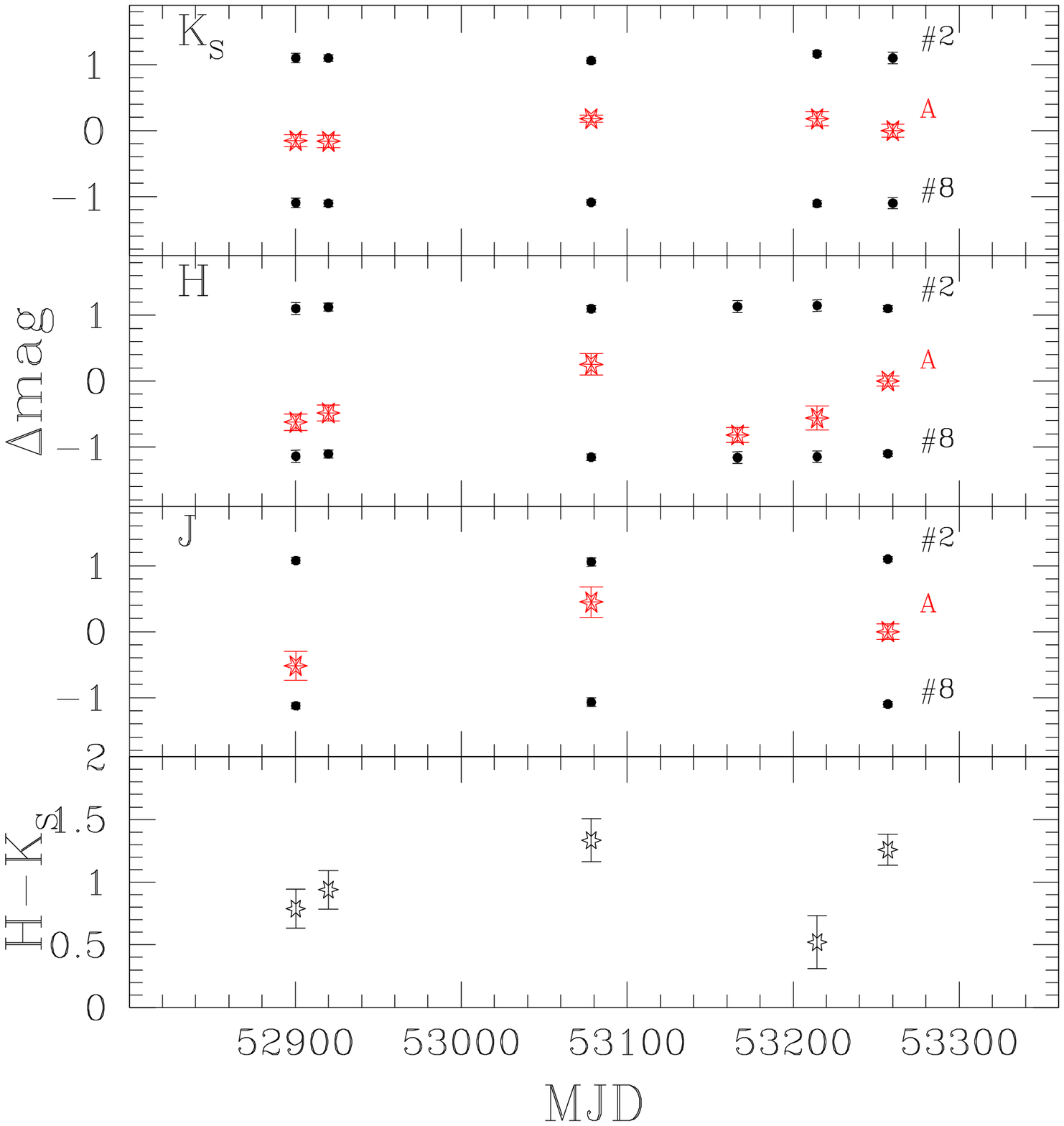} 
 \caption{\xte: {\em Top panel}: X-ray flux (shown in units of
 $10^{-11}$\,erg\,cm$^{-2}$\,s$^{-1}$) history in the 0.5--10\.keV
 energy range (from Gotthelf \& Halpern 2007). {\em Lower panels}:
 from top to bottom: K$_s$, H, and J band relative magnitude variation
 for \xte\ counterpart (red star) and for two nearby, bright
 comparison objects. Magnitude variations are computed relative to the
 September 2004 observation. To better fit in the plot, magnitude
 differences for objects \#2 and \#8 have been shifted by +1.1 and
 -1.1 mag, respectively. Bottom panel shows the H-K$_s$ colour
 variation. Magnitude errors at (1$\sigma$) are shown. These include 
 internal errors, aperture correction errors and zero point errors.}
\label{fig:TimeXTEJ1810} 
\end{figure}

\begin{table*}
\centering
{\scriptsize
\begin{tabular}{cccccccccc}

\hline
\hline
 Source Id. & R.A.(2000) & Dec. (2000)  & 52900 &  52920 & 53078 & 53166 & 53214
& 53257 & 53261 \tabularnewline
\hline
         & \multicolumn{8}{c}{\textbf{K$_s$}} \tabularnewline
\hline
  A & 272.46283 & -19.73110 & 20.81$\pm$0.08 & 20.82$\pm$0.09 & 21.18$\pm$ 0.05
&  & 21.21$\pm$0.10 &  & 21.04$\pm$ 0.09 \tabularnewline 
  1 & 272.46390 & -19.73016 & 13.87$\pm$0.07 & 13.85$\pm$0.04 & 14.26$\pm$ 0.04
&  & 14.01$\pm$0.04 &  & 14.01$\pm$ 0.08 \tabularnewline 
  2 & 272.46391 & -19.73061 & 17.23$\pm$0.07 & 17.25$\pm$0.04 & 17.23$\pm$ 0.04
&  & 17.36$\pm$0.04 &  & 17.30$\pm$ 0.08 \tabularnewline 
  3 & 272.46338 & -19.73070 & 19.09$\pm$0.07 & 19.21$\pm$0.05 & 19.47$\pm$ 0.04
&  & 19.15$\pm$0.05 &  & 19.41$\pm$ 0.08 \tabularnewline
  4 & 272.46335 & -19.73110 & 19.43$\pm$0.07 & 19.37$\pm$0.05 & 19.46$\pm$ 0.04
&  & 19.42$\pm$0.04 &  & 19.47$\pm$ 0.08 \tabularnewline 
  5 & 272.46399 & -19.73098 & 19.52$\pm$0.07 & 19.77$\pm$0.06 & 19.67$\pm$ 0.04
&  & 19.62$\pm$0.04 &  & 19.70$\pm$ 0.08 \tabularnewline 
  6 & 272.46384 & -19.73145 & 18.61$\pm$0.07 & 18.61$\pm$0.04 & 18.60$\pm$ 0.04
&  & 18.66$\pm$0.04 &  & 18.67$\pm$ 0.08 \tabularnewline 
  7 & 272.46327 & -19.72952 & 17.69$\pm$0.07 & 17.70$\pm$0.04 & 17.73$\pm$ 0.04
&  & 17.75$\pm$0.04 &  & 17.75$\pm$ 0.08 \tabularnewline 
  8 & 272.46535 & -19.72897 & 15.48$\pm$0.07 & 15.40$\pm$0.04 & 15.45$\pm$ 0.04
&  & 15.51$\pm$0.04 &  & 15.53$\pm$ 0.08 \tabularnewline 
  9 & 272.46264 & -19.72882 & 15.25$\pm$0.07 & 15.15$\pm$0.04 & 15.23$\pm$ 0.04
&  & 15.27$\pm$0.04 &  & 15.23$\pm$ 0.08 \tabularnewline 
 10 & 272.46082 & -19.73013 & 14.43$\pm$0.07 & 14.19$\pm$0.04 & 14.55$\pm$ 0.04
&  & 14.44$\pm$0.04 &  & 14.29$\pm$ 0.08 \tabularnewline 
 11 & 272.46430 & -19.73252 & 14.83$\pm$0.07 & 14.74$\pm$0.04 & 14.91$\pm$ 0.04
&  & 14.87$\pm$0.04 &  & 14.90$\pm$ 0.08 \tabularnewline 
 12 & 272.46502 & -19.73141 & 16.59$\pm$0.07 & 16.50$\pm$0.04 & 16.51$\pm$ 0.04
&  & 16.63$\pm$0.04 &  & 16.54$\pm$0.08 \tabularnewline 
\hline
 & \multicolumn{8}{c}{\textbf{H}} \tabularnewline
  
\hline
  A  & & &  21.67$\pm$0.12 & 21.81$\pm$0.16 & 22.55$\pm$ 0.07 &  21.48$\pm$ 0.12
&  21.73$\pm$ 0.11 &  22.29$\pm$0.18 &\tabularnewline 
  1  & & &  14.27$\pm$0.05 & 14.20$\pm$0.04 & 14.39$\pm$ 0.04 &  14.24$\pm$ 0.09
&  14.24$\pm$ 0.08 &  14.24$\pm$0.08 &\tabularnewline 
  2  & & &  17.66$\pm$0.05 & 17.68$\pm$0.04 & 17.66$\pm$ 0.04 &  17.69$\pm$ 0.09
&  17.71$\pm$ 0.08 &  17.67$\pm$0.08 &\tabularnewline 
  3  & & &  20.00$\pm$0.06 & 20.11$\pm$0.05 & 20.07$\pm$ 0.04 &  19.90$\pm$ 0.09
&  19.78$\pm$ 0.09 &  20.07$\pm$0.08 &\tabularnewline 
  4  & & &  22.24$\pm$0.14 & 22.05$\pm$0.11 & 22.06$\pm$ 0.09 &  21.66$\pm$ 0.23
&  21.87$\pm$ 0.10 &  22.64$\pm$0.17 &\tabularnewline 
  5  & & &  22.28$\pm$0.28 & 22.74$\pm$0.42 & -               &  21.79$\pm$ 0.21
&  21.68$\pm$ 0.10 &  22.80$\pm$0.11 &\tabularnewline 
  6  & & &  21.17$\pm$0.09 & 21.38$\pm$0.07 & 21.38$\pm$ 0.05 &  21.11$\pm$ 0.10
&  21.15$\pm$ 0.09 &  21.34$\pm$0.09 &\tabularnewline 
  7  & & &  18.16$\pm$0.06 & 18.19$\pm$0.04 & 18.14$\pm$ 0.04 &  18.14$\pm$ 0.09
&  18.15$\pm$ 0.08 &  18.20$\pm$0.08 &\tabularnewline 
  8  & & &  18.95$\pm$0.06 & 19.00$\pm$0.05 & 18.90$\pm$ 0.04 &  18.90$\pm$ 0.09
&  18.93$\pm$ 0.08 &  18.95$\pm$0.08 &\tabularnewline 
  9  & & &  18.42$\pm$0.06 & 18.42$\pm$0.04 & 18.34$\pm$ 0.04 &  18.44$\pm$ 0.09
&  18.41$\pm$ 0.08 &  18.44$\pm$0.08 &\tabularnewline 
 10  & & &  14.62$\pm$0.05 & 14.51$\pm$0.04 & 14.65$\pm$ 0.05 &  14.61$\pm$ 0.09
&  14.61$\pm$0.08  &       -         &\tabularnewline
 11  & & &  15.15$\pm$0.05 & 15.12$\pm$0.04 & 15.19$\pm$ 0.04 &  15.13$\pm$ 0.09
&  15.14$\pm$ 0.08 &  15.12$\pm$0.08 &\tabularnewline 
 12  & & &  16.91$\pm$0.05 & 16.89$\pm$0.04 & 16.83$\pm$ 0.04 &  16.93$\pm$ 0.09
&  16.91$\pm$ 0.08 &  16.90$\pm$0.08 &\tabularnewline 
\hline

 & \multicolumn{8}{c}{\textbf{J}} \tabularnewline
 
\hline
  A & & & 22.92$\pm$ 0.22 &  & 23.89 $\pm$ 0.23 & & &  23.45 $\pm$ 0.12
&\tabularnewline 
  1 & & &      $-$        &  & 15.37 $\pm$ 0.06 & & &  15.04 $\pm$ 0.04
&\tabularnewline 
  2 & & & 18.85$\pm$ 0.05 &  & 18.83 $\pm$ 0.06 & & &  18.88 $\pm$ 0.04
&\tabularnewline  
  3 & & & 21.06$\pm$ 0.06 &  & 21.35 $\pm$ 0.07 & & &  21.35 $\pm$ 0.04
&\tabularnewline  
  7 & & & 19.34$\pm$ 0.05 &  & 19.39 $\pm$ 0.06 & & &  19.36 $\pm$ 0.04
&\tabularnewline  
  9 & & & 18.06$\pm$ 0.04 &  & 18.00 $\pm$ 0.06 & & &  18.01 $\pm$ 0.04
&\tabularnewline 
 10 & & &      $-$        &  & 15.54 $\pm$ 0.06 & & &      $-$         
&\tabularnewline  
 11 & & & 15.98$\pm$ 0.04 &  & 16.01 $\pm$ 0.06 & & &  15.97 $\pm$ 0.04 &
\tabularnewline  
 12 & & & 19.04$\pm$ 0.05 &  & 18.94 $\pm$ 0.06 & & &  18.99 $\pm$ 0.04 &
\tabularnewline 
\hline
\hline
\end{tabular}}
\caption{\xte: From top to bottom: K$_s$, H and J magnitudes for the candidate
IR counterpart of \xte\, and some of the twelve nearby objects, for the 6
different observing runs (see Tab.\ref{tab:obs}). Numbers atop the magnitude
columns give the observations epochs in Modified Julian Days (MJD). 
Magnitude errors are given at 1$\sigma$ and include zero point errors, either
absolute or relative,
that are, for the brightest objects, the dominant error source.}
\label{tab:xteJ}
\label{tab:magXTEJ1810_H}
\label{tab:magXTEJ1810_K}
\label{tab:magXTEJ1810_J}
\end{table*}


\subsection{Data reduction and calibration} 

The    VLT    data    were    pre-reduced   with    the    ESO    NACO
pipeline\footnote{www.eso.org/instruments/naco/}, which is based on 
the package \textit{eclipse}, while the native version of the
\textit{eclipse}\footnote{www.eso.org/eclipse}  package  was used  for
the  Gemini and  CFHT data.   In  all cases,  the image  pre-reduction
produced  a final  co-added, sky-subtracted  and flat-fielded
image for each band. 

Depending on the case, fluxes were computed either through PSF or
aperture photometry using the IRAF\footnote{IRAF is distributed by the
National Optical  Astronomy Observatories,  which are operated  by the
Association  of Universities  for Research  in Astronomy,  Inc., under
cooperative agreement with  the National Science Foundation.}  version
of  the  \textit{daophot}   package  \citep{stet92}.  For  AO
observations we found some instrument-dependent effects which hampered
the  PSF  fitting and  discouraged  the  use  of PSF  photometry.   In
particular, in  the VLT-NACO images the  PSF was found  to be variable
across the field of view, while in the CFHT-AOB images it was found to
feature  an   unusual  profile   with  a  clearly   visible  secondary
diffraction ring.  Both effects are  difficult to account for, even by
using  more sophisticated  and  position-dependent PSF  models. On  the
other hand,  in the Gemini-NIRI  images the PSF  was found to  be more
stable,  although worse  sampled due  to the  larger pixel  size.  For
these  reasons, we  decided  as  a general  strategy  to use  aperture
photometry for our  AO images, i.e. the NACO and AOB  ones, and to use
PSF  photometry  for  the  NIRI  images.  For  the  former,  aperture
photometry was performed using  an aperture or  diameter twice
the  measured  PSF.  Aperture  and PSF photometries have  been
compared   by   applying    aperture   corrections   using   the   IRAF
\textit{daogrow}  algorithm, which  calculates the  aperture correction
extrapolating to an ideally infinite aperture.

For the  VLT images the photometric  calibration was performed
by observing  standard stars  from the \cite{pe98}  catalog, usually
observed under similar conditions (and at about the 
same epoch) as our targets. For
all the NACO runs the zeropoint  uncertainty has been found to be very
small,  i.e.  in  the  0.02  -  0.05  magnitude  range.  The  residual
uncertainty introduced by  the extinction term is almost  wiped out by
the closeness in time and  airmass of target fields and standards. Since it turns out to be $\la 0.01$ magnitudes it has been neglected.

The Gemini images of \xte\ have been at first instance calibrated with the
available standard stars, but the uncertainty in the zero points and
the zero point differences usually found between standards taken on
the same nights suggested us to discard the primary standard
calibration. A photometric calibration was also tried using stars from
the 2MASS \citep{skrutskie06} catalog identified in the target fields.
However, such an on-the-frame calibration suffered from two main
problems.  Firstly, the number of suitable 2MASS stars is often quite
small and the brighter ones were found to be generally saturated in
our images. Furthermore, the 2MASS catalog was built from images with
a 2$\arcsec$ pixel size so that, in many cases, the measured
magnitudes correspond to blended objects which are, instead, resolved
in our higher resolution images.  Thus, given the accuracy of the NACO
zeropoints and the very low r.m.s.  of the photometric solution, we
chose to re-calibrate the Gemini photometry of the \xte\ field on the
NACO one by using, as secondary photometric calibrators, a set of
well-suited field stars in common between the NACO and the NIRI
images.  As a reference, we chose the September 2004 NACO observations
because of their overall better quality. In order to ensure full
consistency among observations performed at different epochs (Tab.\,
\ref{tab:obs}), also the NACO October 2003 and March 2004 photometry
of the \xte\ field was re-calibrated to the September 2004 one.  The
relative photometric uncertainty, based on about 100 stars in common
among the various images is of the order of $\approx$ 0.05 magnitudes
for the NACO-to-NACO registration, and slightly worse for the
Gemini-to-NACO one ($\approx$ 0.08 magnitudes).

For the  CFHT photometry, for which  no NACO images  are available, the
only option was to use the 2MASS catalog, after carefully accounting
for all the caveats mentioned above. 
For the L$^{\prime}$ observation of \rxs\
we have used standard stars from the UKIRT IR standard catalog
\citep{ukirt}, yielding a photometric accuracy of $\sim$ 0.1 magnitudes.

In all cases, astrometric  calibration was performed using 2MASS stars
as  reference, yielding  an  average uncertainty  of $\sim$ 0\farcs2,
after  accounting  for the  intrinsic  astrometric  accuracy of  2MASS
($\sim$ 200 mas) and the  r.m.s. of the astrometric fits. However, due
to the small instrument pixel scales, the latter has been found to be much
smaller  than the  2MASS astrometric  accuracy and  hence it  has been
neglected.

\section{Data analysis and results}

\subsection{\xte}
\label{xte}

Fig.\,\ref{fig:MapXTEJ1810} (left) shows the \xte\ field
(K$_s$ band) observed with NACO on October 2003 (see also Israel et
al. 2004b).  The \xte\ candidate IR counterpart, labeled A, is
coincident with the radio position (RA 18$^{\rm h}$ 09$^{\rm m}$
51$^{\rm s}.$087, DEC -19$^{\circ}$ 43$^{\prime}$ 51\farcs93; error =
0\farcs01; Camilo et al. 2006).  For both the NACO and NIRI observations, photometry was
computed and calibrated as described in the previous section. For each
instrument, the resulting J, H and K$_s$ band catalogues were then
matched and compared with the multi-band ones.  The multi-band
magnitudes of the \xte\ candidate counterpart, all calibrated to the
NACO system, are reported in Tab.\,\ref{tab:xteJ} together with those
of a number of field objects, chosen among those used as a reference
for the photometry cross-calibrations (see previous section).  We note
that the candidate counterpart is not detected in the J and K$_s$-band
Gemini observations of the June 2004 run which were performed in
non-photometric conditions.  Given the short integration time, the overall poor data quality, and the lack of
standard observations, we could only derive tentative estimates on the
limiting magnitudes (3$\sigma$ upper limits of J$>$21.9 and K$_s>$20) which are of no use to constrain the flux
evolution of our source.

The  time-averaged (K$_s$, H-K$_s$)  colour-magnitude diagram  (CMD) of
the  field   is  shown  in   Fig.\,\ref{fig:MapXTEJ1810}  (right).   A
well-defined main  sequence group of stars, probably  tracking a young
star cluster, is recognizable in the left part of the diagram, while a
second red clump  of stars, most likely red  giants or highly absorbed
main sequence  stars, is present  on the right. The  \xte\ counterpart
(green  star) appears much  fainter than  the nearby  comparison stars
(red dots), and its colour indeed  suggests that it might be a peculiar
object.

The variability of the \xte\ IR counterpart is shown in
Fig.\,\ref{fig:TimeXTEJ1810}, together with the X-ray variability of the
X-ray source (see \S\ref{discussion} for details).  While in the
X-rays the source features a clear, almost monotonic flux decay, in
the IR there is no clear trend in the flux evolution, and hence
no correlation with the X-rays. Indeed, the flux seems to vary in an
erratic way in all the three IR bands. In particular, we found that
the overall variability is apparently larger in the H and J bands
(1.07$\pm$0.24 and 0.97$\pm$0.32 magnitudes, respectively) than in the
K$_s$ one (0.40$\pm$0.13 magnitudes).  We note that our detailed
re-analysis of the October 2003 and March 2004 observations of \xte\
yields a K$_s$ band variability slightly smaller (0.36$\pm$0.10) than
that originally reported by Rea et al.  (2004) using the same
observations (0.50$\pm$0.10), although it is still consistent within
the errors.  We attribute this (non significant) difference to the
more accurate relative photometry between the two epochs, performed
using a larger set of secondary photometric calibrators than the one
used in Rea et al. (2004).  The last panel shows the (H-K$_s$) colour
variability of the source.  In general, we note that the source seems
to become redder when its flux decreases, although no clear trend can
be recognized in our sparse photometry.
We need to warn, however,  that any conclusion  on the source  flux and
colour variability must be taken with  due care.  First of all, the
apparent  lack of  correlation between  the variability  in the  H and
K$_s$  bands is  only suggested  by the  July 2004  Gemini  K$_s$ band
observation.  If we take this observation out and we consider only the
co-eval  flux measurements,  the  K$_s$ band  lightcurve becomes  more
consistent with the H (as well as with the J) band one.  More  generally,
although  special  care  was  devoted to  check  the  cross-instrument
calibration (see \S 2.2), it  is possible that our relative photometry
is  affected  by  random  errors.   These may be induced,  e.g.   by
fluctuations  in the atmospheric  conditions, sky  background, seeing,
quality of  the primary calibration frames (dark  and flatfields), and
glitches in the detector  performance.  Of course, while these effects
are marginal  for the relatively bright  stars that we have  used as a
reference  to compute  our  cross-instrument photometric  calibration,
they are indeed significant for  much fainter stars, such as the \xte\
candidate counterpart (K$_s$ $\sim$  21), and can increase the overall
uncertainties on  the measured magnitudes. To  quantify these random
errors we have cross-checked the  photometry of a number of test stars
with brightness comparable to the one of our target. We found that all
test stars show a scatter of $\sim$0.30 magnitudes in J and of
$\sim$0.10 magnitudes  in the H  and K$_s$ bands for  the NACO-to-NACO
registration, while the scatter  increases to $\sim$0.20 magnitudes for
the  Gemini-to-NACO registration.  Hence,  we take  these values  as
representative  of the random  errors of  our relative  photometry. By
adding  them  to the  formal  errors  we find  that  only  the H  band
variability (1.07$\pm$0.26) can be considered formally significant, and
it is consistent with that measured in the J band (0.97$\pm$0.43) and,
marginally, also with the K$_s$ band one (0.40$\pm$0.16).

\begin{table}
\centering
{\scriptsize
\begin{tabular}{lccccc}
\hline
\hline
Id	&	RA ($^{\circ}$) &	DEC  ($^{\circ}$) & J 	&	H
&  Ks 	\tabularnewline	
\hline
 1  &  257.19551 &  -40.14796 &  20.93$\pm$0.09 & 18.60$\pm$0.06 &
17.53$\pm$0.04  \tabularnewline
 2  &  257.19510 &  -40.14815 &  22.23$\pm$0.17 & 19.84$\pm$0.09 &
18.81$\pm$0.05  \tabularnewline
 3  &  257.19560 &  -40.14799 &  22.11$\pm$0.25 & 20.01$\pm$0.12 &
18.85$\pm$0.05  \tabularnewline
 4  &  257.19503 &  -40.14792 &  22.03$\pm$0.17 & 20.19$\pm$0.09 &
19.63$\pm$0.07  \tabularnewline
 5  &  257.19498 &  -40.14800 &      $-$        & 20.50$\pm$0.10 &
19.95$\pm$0.08  \tabularnewline
 6  &  257.19495 &  -40.14800 &      $-$        &     $-$        &
20.03$\pm$0.09  \tabularnewline
 7  &  257.19494 &  -40.14793 &      $-$        & 20.90$\pm$0.11 &
20.25$\pm$0.08  \tabularnewline
 8  &  257.19563 &  -40.14790 &      $-$        &     $-$        &
20.83$\pm$0.10  \tabularnewline
 9  &  257.19502 &  -40.14786 &      $-$        & 21.55$\pm$0.18 &
20.92$\pm$0.11  \tabularnewline
10  &  257.19531 &  -40.14795 &      $-$        &     $-$        &
20.94$\pm$0.11  \tabularnewline
11  &  257.19500 &  -40.14804 &      $-$        &     $-$        &
20.69$\pm$0.10  \tabularnewline
12  &  257.19548 &  -40.14812 &      $-$        &     $-$        &
21.12$\pm$0.19  \tabularnewline
13  &  257.19526 &  -40.14761 &      $-$        &     $-$        &
21.23$\pm$0.14  \tabularnewline
14  &  257.19545 &  -40.14774 &      $-$        &     $-$        &
21.34$\pm$0.21  \tabularnewline
15  &  257.19523 &  -40.14807 &      $-$        &     $-$        &
21.52$\pm$0.17  \tabularnewline
16  &  257.19509 &  -40.14787 &      $-$        &     $-$        &
21.58$\pm$0.12  \tabularnewline
17  &  257.19510 &  -40.14799 &      $-$        &     $-$        &
21.59$\pm$0.16  \tabularnewline
18  &  257.19538 &  -40.14816 &      $-$        & 21.34$\pm$0.15 &
21.66$\pm$0.14  \tabularnewline
19  &  257.19550 &  -40.14813 &      $-$        &     $-$        &
21.74$\pm$0.16  \tabularnewline
20  &  257.19530 &  -40.14792 &      $-$        &     $-$        &
21.88$\pm$0.19  \tabularnewline
\hline
\hline
\end{tabular}}
\caption{\rxs: J,  H, and K$_s$  magnitudes  for all the  objects detected within
1\farcs1
of the source X-ray position, ordered by decreasing K$_s$ magnitude. No object
has been detected in the L$^{\prime}$ band. 
Magnitude errors are given at 1$\sigma$, and include zero point uncertainties.}
\label{tab:magRXSJ1708}
\end{table}

\begin{figure}
\centering
\includegraphics[width=8cm]{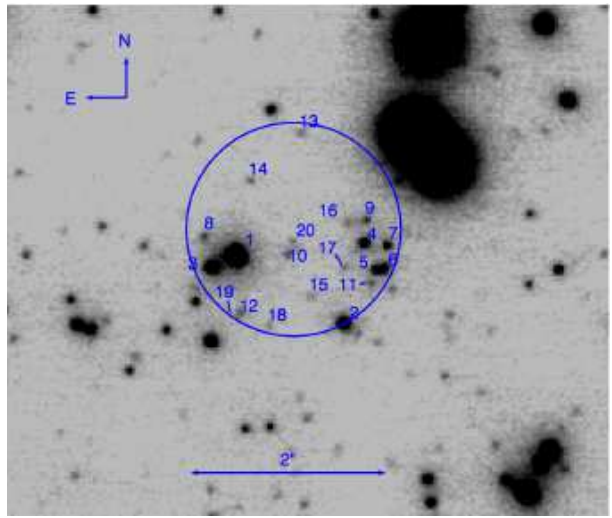}
\caption{\rxs: NACO K$_s$-band image of the source field with the
1\farcs1 radius (99\%  confidence level) X-ray error circle overlaied. Objects
detected at $\ge 3 \sigma$ are labeled.}
\label{fig:MapRXSJ1708}
\end{figure}

\begin{figure*}
\centering
\hbox{
  \includegraphics[width=8cm,height=8cm]{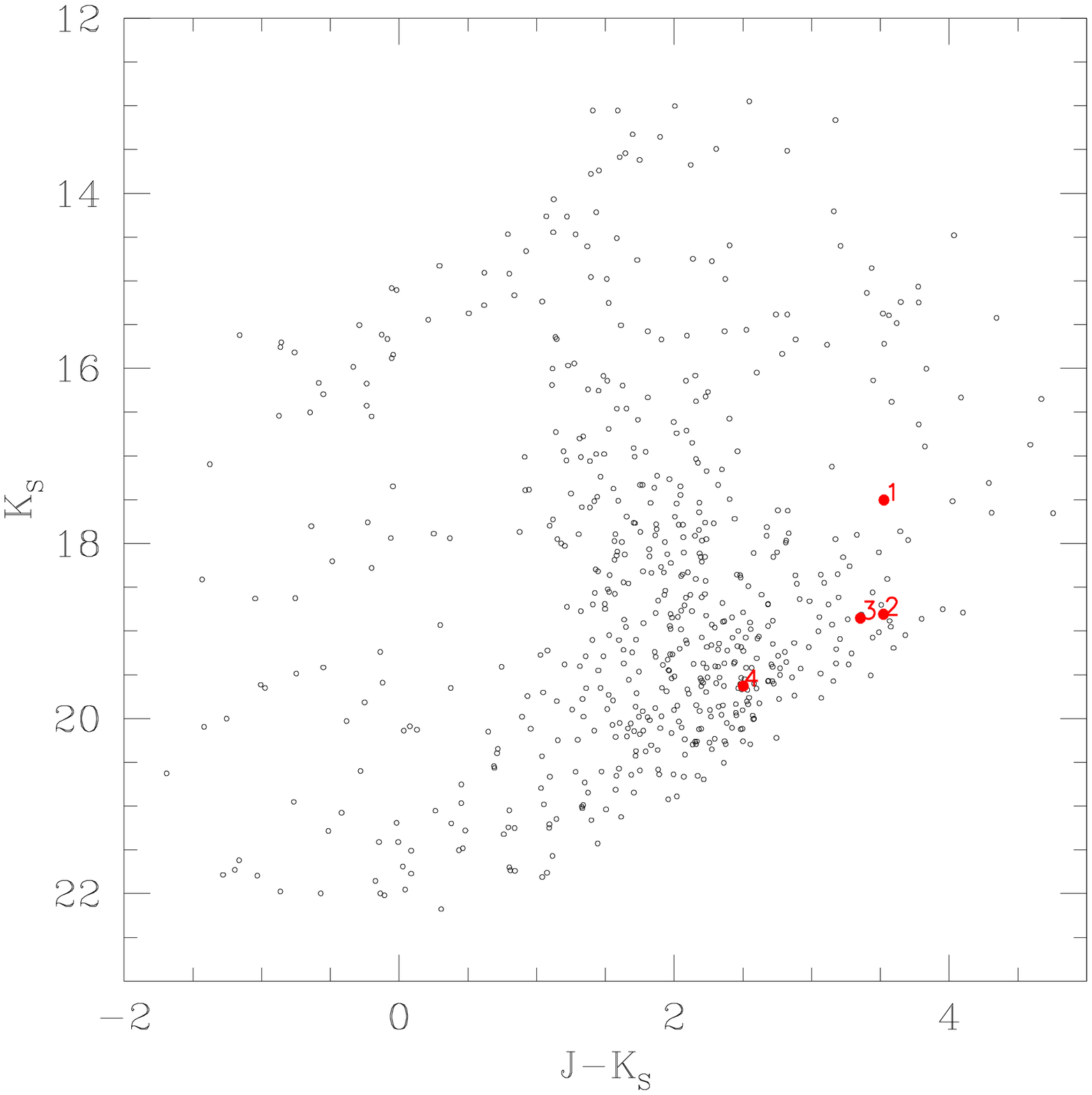}
\hspace{0.5cm}
 \includegraphics[width=8cm,height=8cm]{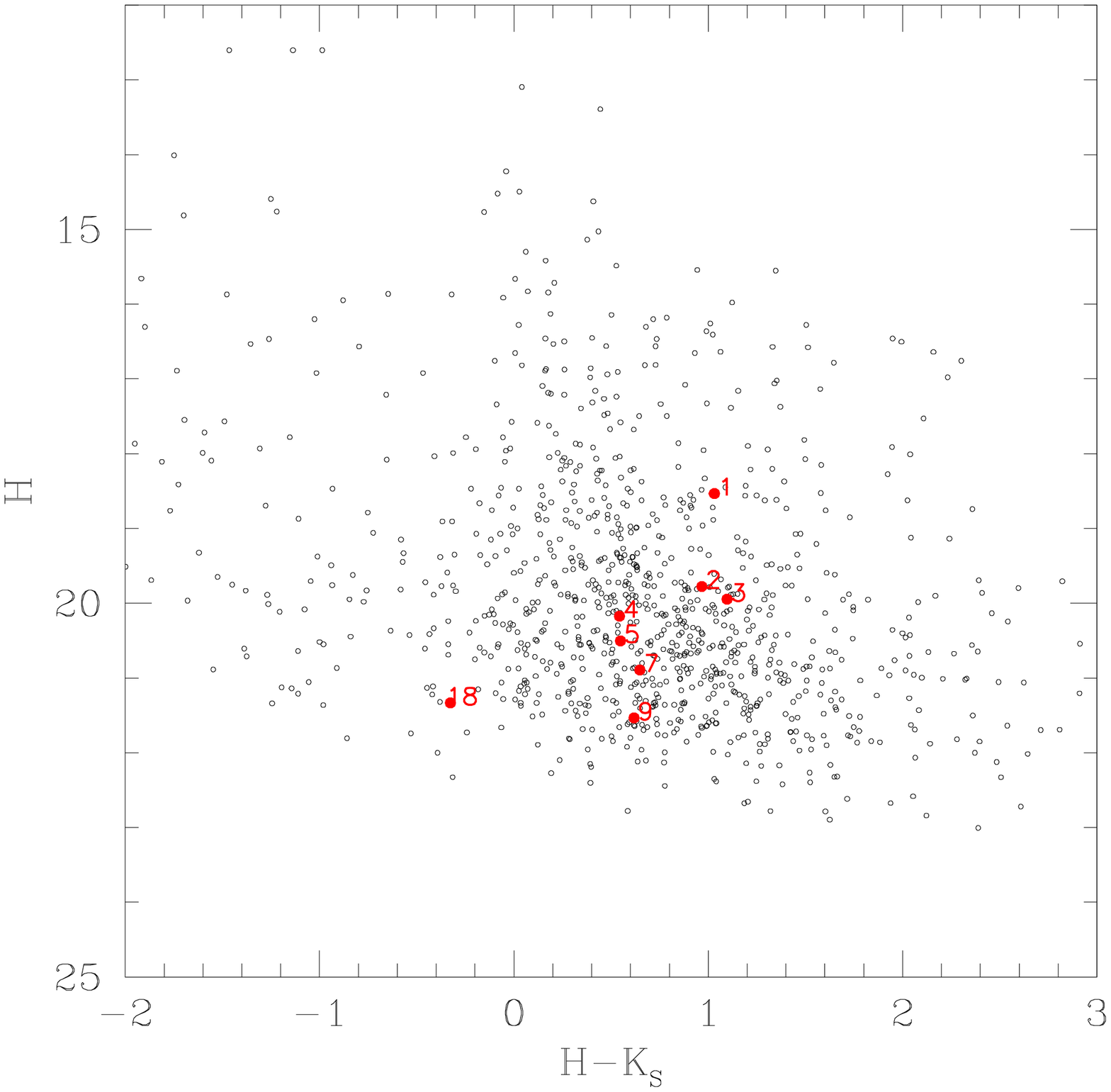} }
 \caption{\rxs: Colour-magnitude diagrams for the objects in the field of view. Sources
detected within or close to 
the X-ray error circle (see Fig.\,\ref{fig:MapRXSJ1708}) are shown in red and
labeled. }
 \label{fig:CmdRXSJ1708}
\end{figure*}

\subsection{\rxs}
\label{rxs}

Fig.\,\ref{fig:MapRXSJ1708}  shows the  NACO $K_s$-band  image  of the
field around  the Chandra position  of \rxs (RA 17$^{\rm  h}$ 08$^{\rm
m}$  46$^{\rm s}.$9, DEC  -40$^{\circ}$ 08$^{\prime}$  52\farcs4). The
size  of the  X-ray error  circle has  been estimated  considering the
uncertainty on  the Chandra  position, that is  0\farcs7 \citep{is03},
combined with the overall accuracy of our astrometric calibration (see
\S2.2), yielding  a final uncertainty  of 1\farcs1 at  99\% confidence
level.  The magnitudes of all the objects detected within the X-ray error
circle are reported in Tab.\,\ref{tab:magRXSJ1708}. No other object is
detected down to the $3 \sigma$ limiting magnitudes of 22.98, 22.20, 22.26
in the J, H, K$_s$ bands, respectively. In the L$^{\prime}$ band no
object is detected within the X-ray error circle down to 17.8 (the
deepest limit ever obtained for an AXPs in this band).  Fig.\,\ref{fig:CmdRXSJ1708} shows the (K$_s$, J-K$_s$) and (H,
H-K$_s$) CMDs of all the objects detected in the field.  The two CMDs show
a fairly scattered sequence. Despite previous claims (Israel et
al. 2003; Safi-Harb \& West 2005; Durant \& van Kerkwijk 2006), none of these objects show significant ($>$3$\sigma$) IR variability to justify a safe identification as the IR counterpart of \rxs\, (see also
\S\ref{discussion}).

\subsection{\kes}
\label{kes}

\begin{figure*}
\centering
\hbox{
 \hspace{0.5cm}
\includegraphics[width=0.49\textwidth]{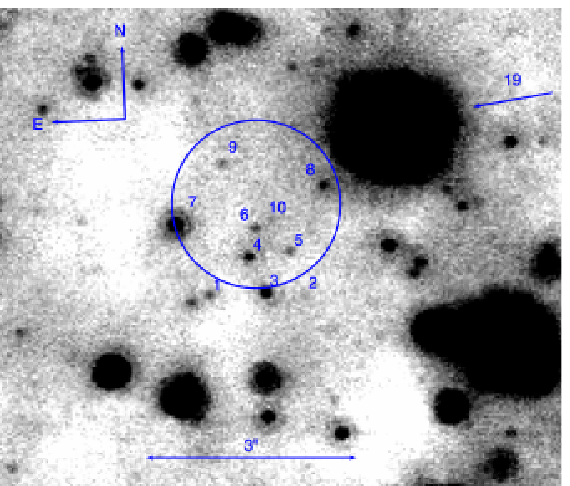}
 \hspace{0.5cm}
\includegraphics[width=8cm,height=8cm]{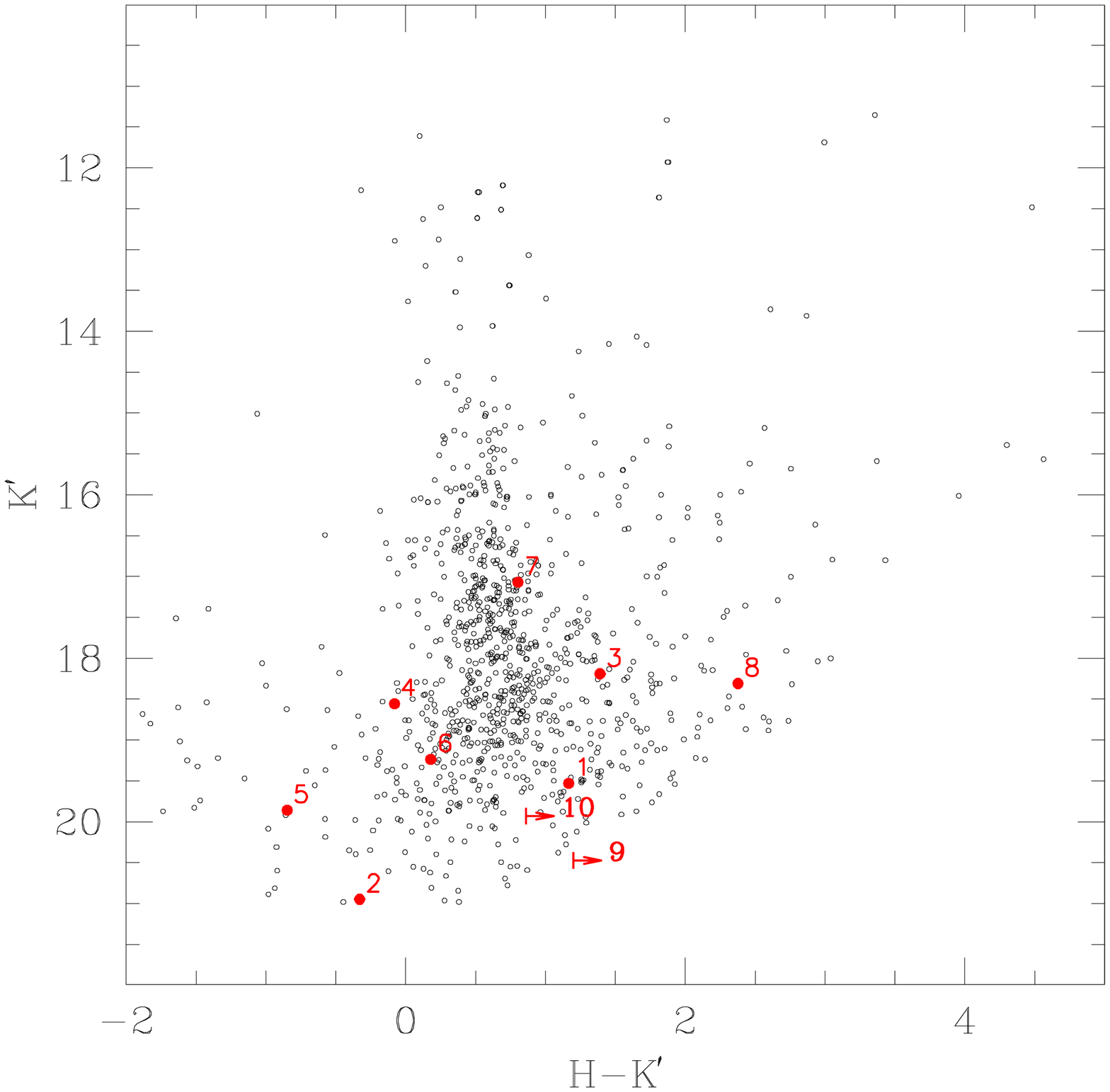}}

\caption{{\em Left panel:} \kes: CFHT K$^{\prime}$-band image of field, with the
1\farcs2 radius (99\%  confidence level) X-ray  error circle overlaied. Objects
detected at 
$\ge 3 \sigma$ are labeled.  Object \#19 of \cite{me01} is labeled as a
reference.
  {\em Right panel:} K$^{\prime}$, H-K$^{\prime}$
  CMD for \kes\ field.  Objects detected within or close to the
X-ray   error circle (left) are shown in red and labeled.}

 \label{fig:Map1E1841}
\label{fig:cmd1E1841}
\end{figure*}

An updated  determination of  the \kes\ position \citep{wac04}
was obtained  with Chandra and gives RA  $18^{\rm h}$  41$^{\rm  m}$ 19$^{\rm
s}$.336,   DEC   -04$^{\circ}$   56$^{\prime}$  10\farcs83,   with   a
3$\sigma$ uncertainty radius of 0\farcs9. However, the latter uncertainty is
based only on five 2MASS reference stars used for the boresight correction (Wachter et al. 2004). We decided to use a conservative nominal Chandra positional
uncertainty of 0\farcs7 at 90\% confidence level.
Fig.\,\ref{fig:Map1E1841} shows a revised finding chart of the field
around  the updated  X-ray position.   The  size of  the X--ray  error
circle, 1\farcs2 (99\% c.l.) accounts  for the overall accuracy of our
astrometric calibration  (see \S2.2).   The magnitudes of  all objects
detected within  or close  to the X-ray  error circle are  reported in
Tab.\,\ref{tab:cat1E1841}. 
No other object has been
detected down to the 3$\sigma$ limiting  magnitudes of 21.54 and 21.00 for
H and  K$^{\prime}$, respectively.  All objects have  been detected in
both   the   H   and   K$^{\prime}$   bands,  except   for   \#9   and
\#10. Fig.\,\ref{fig:cmd1E1841} shows the position of these objects in
the (K$^{\prime}$,H-K$^{\prime}$)  colour-magnitude diagram compared to
other  objects detected in  the field.   We note  that object  \#19 of
\cite{me01}, originally considered a  potential candidate on the basis
of the ROSAT X-ray position, lies  more than two error radii away from
the  updated Chandra  position.   The CMD  shows  a well-defined  main
sequence extending  up to K$^{\prime}$ $\sim$ 12,  with the candidates
occupying  the fainter  region and  stretching over  a broad  range of
colours. 

By comparing our photometric results with those reported in literature for \kes\, (Durant 2005), 
it is evident that our source \#9 (source B in Durant (2005)), shows
a $>$4$\sigma$ variability ($\Delta Ks =$1.30$\pm$0.24).  As a comparison, the average
$\Delta Ks$ and $\Delta H$ between our photometry and that of Durant (2005)
for 9 objects in the field of view is $\sim$0.2 magnitudes. We then tentatively propose this object as the IR counterpart to this AXP.

\subsection{\sgrb}
\label{sgrb}

\begin{figure}
\centering
\hbox{
\includegraphics[width=8cm]{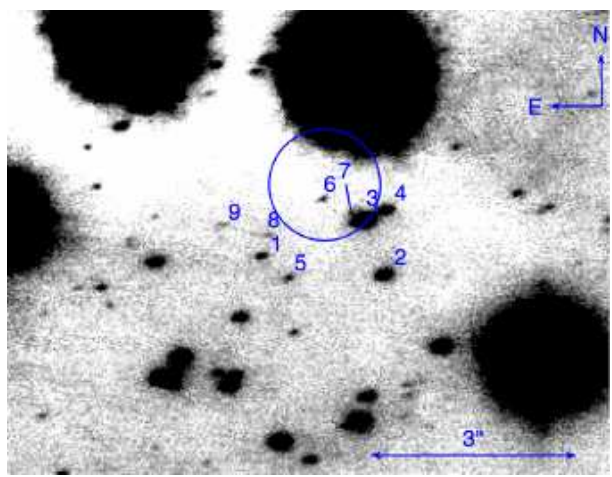}}
\caption{\sgrb. Zoom-in of the area of the K$_s$ band image of 2006 March 31st
around the radio position 
of Frail et al. (1999). The error circle (radius 0\farcs81 at 99\% confidence level) is
shown. Two large ghosts are visible
  atop the position of \sgrb, hiding star E of Kaplan et
  al. (2002). }
\label{fig:MapSGR1900}
\end{figure}

Fig.\,\ref{fig:MapSGR1900}  shows the  NACO  K$_s$ band  image
around the \sgrb\ radio position (RA 19$^{\rm h}$ 07$^{\rm m}$
14$^{\rm s}$.33, DEC +09$^{\circ}$ 19$^{\prime}$ 20\farcs1, with a 1
$\sigma$ uncertainty of 0\farcs15 in each coordinate; Frail et al. 1999).  Our astrometry
yields an overall 3$\sigma$ position uncertainty of 0\farcs81, which
compares well with the one of 0\farcs79 (99\% confidence level) quoted
by \cite{ka02}.  We note that the PSFs of field objects are slightly
elongated in the NACO images, and one of the objects detected within
the radio error circle (our \#3 and object C of Kaplan et al.~(2002)) looks
apparently extended.  However, its PSF is consistent with that
measured for most field objects.   In this case, photometric
calibration was performed using as reference the magnitudes of a few
objects detected around the target position \citep{ka02,fr99} (see Fig.\,\ref{fig:MapSGR1900}). The three new
objects detected in our NACO images are also marked (\# 7, 8 and 9).  In 
particular, object \#7 falls very close to the position of \#3, which prompted us to use PSF photometry rather
than aperture photometry as done, instead, for the other NACO
observations.  The magnitudes of all the detected objects are listed in Tab.\,
\ref{tab:catSGR1900}.No other object is detected close or within the
radio position error circle down to a 3$\sigma$ limiting magnitude of
K$_s$ = 22.60 and 22.25 for the March 31 and for the July 20
observations, respectively.  We note that among our newly detected
objects, \#7 (K$_s$ $\sim$ 19.7) is the only new source within the 99\% error
circle, and it displays some evidence of variability ($\Delta$~K$_s=$0.47$\pm$0.11) between the two epochs.  Unfortunately,
its proximity to object \#3 prevents a better determination of the
actual magnitude variation, even through PSF photometry. As we did in
\S  3.1, we evaluated  the effects  of random  errors on  our relative
photometry  by measuring  the magnitude  scatter of  a number  of test
stars of brightness comparable to object \#7 and selected close to the
field center  in order not to  introduce biases in  our PSF photometry
due to  the position-dependent NACO PSF.   We found that  for the test
stars the scatter is $\approx  0.1$ magnitudes. Even adding the random
error  we found  that  the measured  variation  is still  significant,
although only at the $\sim  3 \sigma$ level.  Based on variability, we
then propose object \#7 as a tentative IR candidate counterpart to
\sgrb.

\section{Discussion}
\label{discussion}

Before discussing our results, we would like to clarify here how we define a reliable identification of an IR counterpart to an AXP or SGR. All of the current well defined counterparts have
been established either through a very accurate positional coincidence
(no other candidates in the positional error circle) or by the detection of significant  IR variability.  In some cases, what are usually defined (not very clearly) as "strange" colours with respect to the
other stars in the field of view has been used as a method to claim the detection of a counterpart. However, we prefer not to adopt this "colour" criterion because we find it rather misleading. In fact, i) almost all counterparts reliably identified through IR variability detection or positional coincidence have colours not much at variance with some of the their field stars, and ii) our current lack of knowledge of the the exact physics behind the IR emission of AXPs and SGRs does not allow any prediction on their IR spectra. Another criterion, often used to propose or strengthen an identification is the X/IR flux
ratio. We believe this method rather rough and not constraining at all for the following reasons: i) AXPs and SGRs are variable both in the X-ray and IR bands, and we do not have yet a clear idea on the connection between  these two bands, ii) there are no theoretical predictions with respect to the X/IR ratio that an AXP or SGR is expected to have, and iii) with the exception of their IR variability, the IR counterpart to these neutron stars are, at our current knowledge, consistent with the IR emission of very low mass stars, which if laying in the source error circle will then result in the same X/IR ratio. This is why as long as significant variability is not detected, if more
than one faint IR object is present in the positional circle, any proposed IR conterpart needs a further confirmation.

\subsection{\xte\, IR variability}

IR observations with the VLT--NACO camera of \xte\, performed
after the outburst, unveiled its IR counterpart
\citep{is04b}. Follow-up observations performed six months later
detected IR variability from the candidate counterpart\citep{re04},
with a flux decrease by about a factor of 2. Simultaneous X--ray
observations over the same period revealed a similar X-ray flux
decrease between the two epochs (Rea et al.  2004; Gotthelf et al.
2004).  This suggested that the
correlated IR/X-ray variability is a characteristic of this source,
as expected if the IR emission were dominated by the reprocessing of
the X-ray emission from a fossil disk around the magnetar (Perna et
al.  2000; Perna \& Hernquist 2000). However, our recent IR
observations show that such correlation is not as obvious as thought
before (see also Camilo et al.~2007b). While the IR flux is indeed
variable, it does not follow the same monotonic decrease of the X-ray
flux observed in the post-outburst phase (see Fig.\ref{fig:TimeXTEJ1810} 
and Gotthelf \& Halpern 2007). Instead, the IR variability appears more 
similar to the one observed in radio band (Camilo et al. 2006, 2007c). 
Unfortunately, the IR observations are
too sparse for any firm conclusion to be drawn regarding a possible
connection between the radio and the IR variability. Furthermore, we
note that the onset of the radio emission (January 2005; Camilo et
al. 2006) occurred after our IR observations.

In the fossil disk scenario, the \xte\, IR variability can be hardly
explained if the IR emission were dominated by reprocessing of the
X-ray radiation from the pulsar. In this case in fact the IR flux
should decrease as the X-ray flux drops (see Rea et al. 2004 for
details), and no increase in any of the IR bands would be
expected. If, on the other hand, the IR flux were dominated by the
disk emissivity resulting from viscous dissipation (which could be
the case if the disc inner radius is farther out), then flux
variability would be the result of a variation in the disk mass inflow
rate.  Given the non-correlated IR and X-ray variability we observe,
this latter scenario could be more viable. However, while a
variability in the IR emission is easily predictable as resulting from
a mass inflow rate variation, this should take place similarly in all
the IR bands. As noted in \S\ref{xte}, we see a hint for an IR-band
dependent variability, although no significant conclusion can be drawn
on this aspect yet.  

Within the magnetars scenario, a few different models have been
proposed for the production of the IR radiation (e.g. Heyl \&
Hernquist 2005; Beloborodov \& Thompson 2007), but none of them makes
(yet) specific predictions for the IR variability that would permit a
direct comparison with our data.

IR variability has been observed for other magnetars, i.e. \ee\,
(Israel et al. 2002; Tam et al. 2007), \uu (Hulleman et al. 2004),
\sgra\, (Israel et al. 2005), and \ea\, (Tam et al. 2004). 
Unfortunately, with the current data no firm conclusion can be drawn
yet about the IR variability in connection with variabilities in other
observing bands, except for \ea\ and \sgra\ (Tam et al. 2004; Israel et al
2005), where a correlation with the source bursting activity and X-ray 
flux enhancement has been detected.

\begin{table}
\centering
{\scriptsize
\begin{tabular}{lcccc}
\hline
\hline
Id. & RA  ($^{\circ}$) & DEC ($^{\circ}$) &   H & K$^{\prime}$  \tabularnewline
\hline
   1 & 280.330811 &  -4.936699 & 20.70$\pm$ 0.31  & 19.53$\pm$ 0.15 
\tabularnewline
   2 & 280.330444 &  -4.936720 & 20.62$\pm$ 0.29  & 20.95$\pm$ 0.49 
\tabularnewline
   3 & 280.330597 &  -4.936695 & 19.58$\pm$ 0.10  & 18.19$\pm$ 0.03 
\tabularnewline
   4 & 280.330658 &  -4.936555 & 18.48$\pm$ 0.03  & 18.56$\pm$ 0.05 
\tabularnewline
   5 & 280.330475 &  -4.936538 & 19.01$\pm$ 0.06  & 19.86$\pm$ 0.18 
\tabularnewline
   6 & 280.330627 &  -4.936442 & 19.42$\pm$ 0.08  & 19.24$\pm$ 0.11 
\tabularnewline
   7 & 280.330933 &  -4.936414 & 17.88$\pm$ 0.02  & 17.07$\pm$ 0.02 
\tabularnewline
   8 & 280.330353 &  -4.936280 & 20.69$\pm$ 0.29  & 18.31$\pm$ 0.04 
\tabularnewline
   9 & 280.330760 &  -4.936186 & $>$21.54          & 20.47 $\pm$ 0.28 
\tabularnewline
  10 & 280.330670 &  -4.936334 & $>$21.54          & 19.93 $\pm$ 0.16 
\tabularnewline
\hline
\hline
\end{tabular}}
\caption{\kes: H and K$^{\prime}$ magnitudes for all the objects detected within
1\farcs1 of the source X--ray position. Magnitude errors are given at 1$\sigma$
confidence level.}
\label{tab:cat1E1841}
\end{table}


\subsection{Other sources: \kes, \sgrb\, and \rxs}

No IR counterpart had been claimed so far for \kes\, and \sgrb, likely because of the crowding of the field
in which these magnetars happens to lay.   By comparing
our photometric results with those reported in literature for \kes, it is evident that there is one object, our source \#9 (source B in Durant (2005)), with magnitudes similar to the other AXP counterparts, and showing a $>$4$\sigma$ variability. Furthermore, comparing our two observations of \sgrb\, we found a $\sim$3$\sigma$ variability in source \#7 (not detected by Kaplan et al.  (2002) because it was too faint for their limiting magnitudes). We therefore consider these objects as possible IR counterparts to these two neutron stars. 

The new  deep observation of \rxs\,  show 20 sources within and in the vicinity of 
the 99\% X-ray error circle (see Fig. \ref{fig:MapRXSJ1708}). None of these objects showed significant variability with respect to other observations in the literature, which could help identifying them as reliable magnetar counterparts.  

Note that the magnitudes of both the previous candidates (our \#1 and \#3;  Israel et al. 2003, Safi-Harb \& West 2005, Durant \& van~Kerkwijk 2006) would make \rxs\, much brighter in the IR band than any other  AXP (see e.g. Israel et al. 2004a). Furthermore, all the other fainter candidates we report here, laying within the source 
 positional errors (see Fig.\,3), have brightness in better agreement with the IR
emission properties of AXPs than both objects \#1 and \#3.  \cite{du06} proposed source \#3 as a possible candidate on the basis of IR variability, the strange colours and the resulting X/IR ratio of this object.  We believe that given the large number of faint IR sources laying within the \rxs\, positional error circle, and the rather bright magnitudes of source \#3,   neither the  $\sim 2.5\sigma$ IR variability, nor its colours (see Fig. \ref{fig:CmdRXSJ1708}) or its X/IR ratio  can reliably tight this object to the AXP (see also above). We would like to stress that we can not exclude any of these candidates, but that a firm identification of an IR counterpart to \rxs\, is far from being a settled issue.

This AXP has been recently observed to have a variable X-ray emission
(Rea et al. 2005; 2007b; Campana et al. 2007), which, if correlated
somehow with the IR as for \ea\,(Tam et al. 2004), would imply a
variable IR counterpart. In particular, the X-ray variability observed for
\rxs\, in X-ray observations sparse over several years was of the order of
50\%, which would roughly imply (if X-ray and IR are directly
correlated) an IR variability of $\sim0.5$ magnitudes, not detected from any
of the objects near the position of \rxs.  However, the lack of simultaneous IR and X-ray observations of this source prevent us from drawing any firm conclusions about possible correlated variability.

\begin{table}
\centering
{\scriptsize
\begin{tabular}{lccccc}
\hline
\hline
  Id.  &   RA  ($^{\circ}$)    &      DEC  ($^{\circ}$)   & K(2006.03.31)
& K(2006.07.20)  \tabularnewline\hline
   1   &  286.81000 &   9.32198 & 19.36$\pm$0.09 & 19.39$\pm$0.06 \tabularnewline
   2   &  286.80949 &   9.32190 & 18.14$\pm$0.07 & 18.05$\pm$0.05  \tabularnewline
   3   &  286.80957 &   9.32212 & 17.31$\pm$0.07 & 17.25$\pm$0.05  \tabularnewline
   4   &  286.80948 &   9.32216 & 18.42$\pm$0.07 & 18.48$\pm$0.05  \tabularnewline
   5   &  286.80989 &   9.32189 & 20.10$\pm$0.11 & 19.96$\pm$0.08  \tabularnewline
   6   &  286.80974 &   9.32220 & 20.73$\pm$0.26 & 20.85$\pm$0.22 
\tabularnewline
   7      &  286.80963 &   9.32215 & 19.21$\pm$0.08 & 19.68$\pm$0.08 
\tabularnewline
   8      &  286.80997 &   9.32206 & 20.86$\pm$0.19 & 20.55$\pm$0.13 
\tabularnewline
   9      &  286.81016 &   9.32210 & 20.59$\pm$0.20 & 20.46$\pm$0.15 
\tabularnewline
\hline
\hline
\end{tabular}}
\caption{\sgrb: K magnitudes of the March and July 2006 observations, for 
  all objects within 1\farcs6 from the position (see text for details).  Errors on the magnitudes are given at 1$\sigma$ confidence level.}
\label{tab:catSGR1900}
\end{table}

\section{Conclusions}

The IR counterpart to \xte\, is confirmed to be variable in time (as previosuly proposed by Rea et al. (2004)). However, despite previous claims, its variability might be more similar to its radio behavior than to its X--ray variability,
which appears to decay smoothly after the outburst. Simultaneous IR and radio observations 
of this AXP are needed to reliably assess this possibility.

To date, IR counterparts to AXPs and SGRs have been confirmed only for
\xte\,(Israel et al. 2004b), \ee\,(Wang \& Chakrabarty 2002; Israel et al.
2002), \ea\,(Hulleman et al. 2001), \uu (Hulleman et al. 2000) and \sgra (Israel et al. 2005; Kosugi et al. 2005). 
For the AXP \kes\, and \sgrb, we propose here two possible candidates based on the detection of IR variability.  For all the remaining AXPs and SGRs we still miss a candidate or confirmed IR counterpart (see e.g. Wachter et al. 2004; Durant \& van~Kerkwijk 2005, 2007; Muno et al. 2006; Gelfand \& Gaensler 2007).

Ours and others results on optical and IR observations of AXPs and
SGRs, show that up to now we are far from having an overall picture of
the optical and IR behavior of these neutron stars. Further
observations, possibly simultaneously on a wide energy range, are
needed to refine current theoretical models and shed light on AXPs' and SGRs'  optical and IR emission mechanisms.

\begin{acknowledgements}

Based on observations collected at the European Southern Observatory,
Paranal, Chile,  under program ID  071.D-0503, 072.D-0297, 073.D-0381,
076.D-0383, 277.D-5005,  at the Canada-France-Hawaii  Telescope (CFHT)
which  is operated  by the  National Research  Council of  Canada, the
Institut National des Sciences de  l'Univers of the Centre National de
la Recherche Scientifique of France, and the University of Hawaii, and
at the  Gemini Observatory,  which is operated  by the  Association of
Universities  for  Research in  Astronomy  Inc.,  under a  cooperative
agreement  with the  NSF  on  behalf of  the  Gemini partnership:  the
National   Science  Foundation  (United   States),  the   Science  and
Technology Facilities Council  (United Kingdom), the National Research
Council  (Canada), CONICYT  (Chile), the  Australian  Research Council
(Australia), CNPq  (Brazil) and CONICET  (Argentina). This publication
makes use of  data products from the Two Micron  All Sky Survey, which
is a joint project of the University of Massachusetts and the Infrared
Processing  and Analysis  Center/California  Institute of  Technology,
funded by  the National Aeronautics  and Space Administration  and the
National Science  Foundation. We thank  the Yepun (UT4) and  NACO team
for their constant  help on the observation optimization  during the 3
years covered  by this  program, and the referee for his/her very careful reading of the paper
and constructive report . VT and  GLI acknowledge  support from
MIUR  funds.  NR  is  supported  by  NWO  {\tt  Veni}  Fellowship.  SZ
acknowledges   STFC  (ex-PPARC)  for   support  through   an  Advanced
Fellowship.

\end{acknowledgements}

\end{document}